\documentclass[twocolumn,showpacs,preprintnumbers,amsmath,amssymb,prl]{revtex4}

 \def\XXint#1#2#3{{\setbox0=\hbox{$#1{#2#3}{\int}$}
     \vcenter{\hbox{$#2#3$}}\kern-.5\wd0}}

\def\fakebold#1{\relax\ifvmode\leavevmode\fi%
\ifmmode%
\setbox0=\hbox{$#1$}%
\else%
\setbox0=\hbox{#1}%
\fi%
\kern-.02em\copy0 \kern-\wd0%
\kern .04em\copy0 \kern-\wd0%
\kern-.0125em\raise.02em\box0%
}%

\usepackage{graphicx} \usepackage{dcolumn} \usepackage{bm}

\begin{document}



\title{Spin Polarization Dependence of Carrier Effective Mass in
  Semiconductor Structures: Spintronic Effective Mass}

\author{Ying Zhang} 
\author{S. Das Sarma} 
\affiliation{Condensed
  Matter Theory Center, Department of Physics, University of Maryland,
  College Park, MD 20742-4111}

\date{\today}

\begin{abstract}
We introduce the concept of a spintronic effective mass for
spin-polarized carriers in semiconductor structures which arises from
the strong spin-polarization dependence of the renormalized effective
mass in an interacting spin-polarized electron system. The
majority-spin many-body effective mass renormalization differs by more
than a factor of 2 at $r_s = 5$ between the unpolarized and the fully
polarized two-dimensional system whereas the polarization dependence
($\sim 15\%$) is more modest in three-dimension around metallic
densities ($r_s \sim 5$).  The spin polarization dependence of carrier
effective mass is of significance in various spintronic applications.
\end{abstract}


\maketitle


Spintronics~\cite{zutic} involves extensive manipulation of spin
polarized carriers in semiconductors. We show in this Letter that such
a spin-polarized semiconductor carrier system would have a new kind of
carrier effective mass, the `spintronic effective mass' associated
with it. This spintronic effective mass will depend crucially on both
carrier density and carrier spin polarization.

\begin{figure}[htbp]
\centering \includegraphics[width=2in]{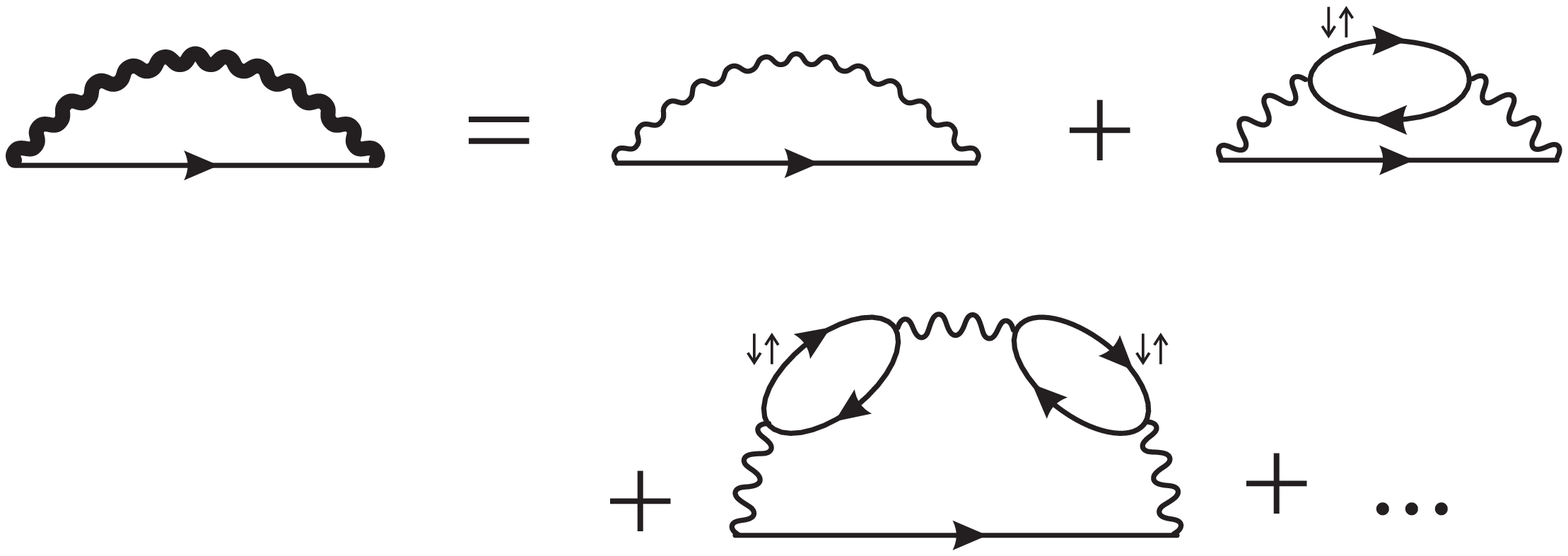}
  \caption{Feynman diagram for self-energy: Solid lines denote the 
    free electron Green's function and the wiggly lines the bare
    Coulomb interaction. At each vertex there is a conserved spin
    index which has to be explicitly accounted for in calculating the
    spin-polarization dependent self-energy.}
\label{fig:feynman}
\end{figure}

Recent experimental measurements of \cite{vitkalov, shashkin, pudalov,
  tutuc, tan} of various Fermi liquid parameters, such as the
effective mass and the spin susceptibility, in two dimensional (2D)
carrier (both electron and hole) systems confined in semiconductor
structures have vigorously renewed interest in one of the oldest
problems \cite{gellman, galitski} of quantum many body theory, namely,
the density dependence of quasiparticle many body renormalization in
interacting electron systems. The quasiparticle effective mass,
$m^*(r_s)$, depends on the interaction parameter $r_s$, the so-called
Wigner-Seitz radius, which is the dimensionless inter-particle
separation measured in the units of the effective Bohr radius: $r_s
\propto n^{-1/2} (n^{-1/3})$ in 2D (3D), where $n$ is the respective
2D (3D) density.  (In 2D systems there is also a non-universal
correction arising from the finite width of the quasi-2D layer in the
quantization direction, which is a conceptually simple form-factor
effect.) In general, $m^*(r_s)$ increases with increasing $r_s$ (i.e.
with decreasing density), except at very small $r_s$, and for 2D
systems of current experimental interest, the many-body
renormalization could be by as much as a factor of $2-3$ at
experimentally relevant densities ($r_s \approx 5 - 10$). In this
Letter we discuss another fundamental new aspect of the quasiparticle
effective mass renormalization which, while being quite important
quantitatively (particularly in 2D semiconductor systems of current
interest), has received relatively minor attention. This is the
dependence, $m^*(r_s, \zeta)$, of the quasiparticle effective mass on
the spin-polarization parameter $\zeta$ ($\equiv |n_{\uparrow} -
n_{\downarrow}|/n$, where $n \equiv n_{\uparrow} + n_{\downarrow}$ and
$n_{\uparrow}$, $n_{\downarrow}$ being the spin-polarized carrier
density). The fact that many-body effects must depend non-trivially on
the spin-polarization parameter $\zeta$ (in addition to the density
parameter $r_s$) is obvious -- for example, the completely
spin-polarized $\zeta = 1$ system has a factor of 2 lower density of
states at $E_F$ (and a concomitantly larger, by $2$ in 2D, Fermi
energy), leading to substantially different many-body renormalization
than the corresponding spin unpolarized ($\zeta = 0$) paramagnetic
system. We provide in this Letter the first complete calculation of
the 2D and the 3D quasiparticle effective mass renormalization $m^*
(r_s, \zeta)$ as a function of both $r_s$ and $\zeta$ within the
leading-order single-loop self-energy expansion
(Fig.~\ref{fig:feynman}) in the dynamically screened Coulomb
interaction (i.e. the infinite series of ring diagrams approximation).
Our calculated effective mass renormalization manifests nontrivial
dependence on the spin polarization parameter $\zeta$, which clearly
needs to be incorporated in understanding the existing experimental
data.

Our theory for the spin-polarization dependence of quasiparticle
effective mass renormalization in interacting electron systems is
motivated not only by fundamental many-body considerations, but also
by practical and urgent experimental needs. In particular, the recent
measurements~\cite{vitkalov, shashkin, pudalov, tutuc, tan} of 2D
effective mass in semiconductor structures invariably involve the
application of an external magnetic field (either parallel or
perpendicular to the 2D layer and often both) which polarizes the
carrier spin. This makes the interpretation of the measured effective
mass as only a density-dependent Fermi liquid parameter $m^*(r_s)$,
perhaps with the appropriate quasi-2D layer width corrections, highly
conceptually suspect since the effective mass in such a spin-polarized
system, $m^*(r_s, \zeta)$, should depend strongly on both density
(i.e.~$r_s$) and spin-polarization (i.e.~$\zeta$). (The precise
magnitude of the spin-polarization factor $\zeta$ is often not
independently known, making an interpretation of the 2D effective mass
measurement difficult.) In addition to the direct connection to the
experimental 2D effective mass measurements, whose analyses must now
be re-examined in light of our theoretical results, our calculated
spin-polarization dependence of quasiparticle effective mass should
also be of considerable significance to the fledgling subject of
spintronics~\cite{zutic} where spin-polarized carriers are manipulated
in semiconductor structures for logic and memory microelectronics
applications. In spintronics applications, e.g. spin Hall
effect~\cite{kato, wunderlich} and spin transistors~\cite{datta,
  zutic2}, the carriers are often spin-polarized, and therefore the
carrier effective mass would depend non-trivially on the
spin-polarization parameter $\zeta$ as predicted in our work. Such a
spin polarization dependence of the carrier effective mass has so far
not been taken into account in the spintronics literature to the best
of our knowledge, but could turn out to be important in understanding
experimental data and device modeling of spintronic systems.

The many-body self-energy Feynman diagrams we calculate for evaluating
the quasi-particle effective mass are shown in Fig.~\ref{fig:feynman}.
This is the single-loop dynamical screening approximation (i.e. the
infinite series of ring diagrams) for the self-energy, generalized to
the 2-component spin-polarized situation. The spin up (down)
quasiparticle self-energy with momentum ${\bf k}$ and frequency
$\omega$ in a polarized electron system can be written within our
approximation as (see, e.g. \cite{rice})
\begin{eqnarray}
\label{eq:E0}
\Sigma^{\uparrow(\downarrow)}({\bf k}, \omega) 
&=& - \int {d^d q d \nu \over i (2 \pi)^{d+1} } 
{v_q \over \epsilon({\bf q}, \nu)} \nonumber\\
&& ~~~~~~~~ \times
G_0^{\uparrow(\downarrow)} ({\bf q} + {\bf k}, \nu + \omega), 
\end{eqnarray}
where $\hbar$ is always chosen to be 1, $v_q$ is the bare Coulomb
interaction between electrons with $v_q = 2 \pi e^2 /q$ in 2D and $v_q
= 4 \pi e^2 /q^2$ in 3D,
\begin{equation}
\label{eq:G0}
G_0^{\uparrow(\downarrow)}({\bf k}, \omega) =
{1 - n_F(\xi_{\bf k}^{\uparrow(\downarrow)}) 
\over \omega - \xi_{\bf k}^{\uparrow(\downarrow)} + i\eta } +
{n_F(\xi_{\bf k}^{\uparrow(\downarrow)}) \over \omega 
- \xi_{\bf k}^{\uparrow(\downarrow)} - i\eta } 
\end{equation}
is the Green's function for free spin up (down) electrons with the
noninteracting energy dispersion $\xi_{\bf k}^{\uparrow(\downarrow)} =
(k^2 - k_F^{2~\uparrow(\downarrow)})/2m$ with $m$ being the bare band
mass, and $\epsilon({\bf k}, \omega)$ is the dynamic dielectric
function. Here $k_F^{\uparrow(\downarrow)}$ is the Fermi momentum for
the spin up (down) electrons. $k_F^{\uparrow(\downarrow)} = k_F
\sqrt{1 \pm \zeta}$ for 2D and $k_F^{\uparrow(\downarrow)} = k_F (1
\pm \zeta)^{1/3}$ for 3D, and $k_F$ is the Fermi momentum for the
unpolarized state.  We use $\eta$ to denote an infinitesimal positive
number, and $n_F(x)$ the Fermi function. At zero temperature, $n_F(x)
=1$ when $x \le 0$ and $0$ otherwise. Within our approximation, the
dynamical screening is done by the infinite series of ring diagrams,
and we have
\begin{equation}
\label{eq:epsilon}
\epsilon({\bf k}, \omega) = 1 
- v_q [\Pi^\uparrow ({\bf k}, \omega) 
+ \Pi^\downarrow ({\bf k}, \omega) ] 
\end{equation}
with $\Pi^{\uparrow(\downarrow)}({\bf k}, \omega)$ the noninteracting
electronic polarizability (i.e. the bare bubble in
Fig.~\ref{fig:feynman}):
\begin{equation}
\label{eq:bubble}
\Pi^{\uparrow(\downarrow)}({\bf k}, \omega) 
= \int {d^2 q \over (2 \pi)^2} 
{n_F(\xi_{\bf q}^{\uparrow(\downarrow)}) 
- n_F(\xi_{{\bf q} + {\bf k}}^{\uparrow(\downarrow)}) 
\over \xi_{\bf q}^{\uparrow(\downarrow)} 
- \xi_{{\bf q} + {\bf k}}^{\uparrow(\downarrow)} + \omega},
\end{equation}

We note that this single-loop self-energy of Fig.~\ref{fig:feynman}
becomes asymptotically exact in the high density $r_s \to 0$ limit,
but is known to give reasonable results in the low-density $r_s > 1$
regime also~\cite{rice, mass}. The reason for this approximate
validity of the single-loop self-energy well in to the strong coupling
regime is the fact that the dynamical screening expansion is {\em not}
a series expansion in $r_s$, but is a self-consistent mean-field
approximation where the effective expansion parameter is akin to $r_s
/ (C + r_s)$ with $C \gg 1$. Once the real part of the carrier
self-energy, ${\rm Re}~\Sigma^{\uparrow(\downarrow)}(k, \omega)$, is
obtained, the effective mass is calculated by the on-shell
quasiparticle approximation~\cite{rice, mass}:
\begin{equation}
\label{eq:onshell}
{m \over m^*_{\uparrow(\downarrow)}} = 1 
+ {m  \over k_F^{\uparrow(\downarrow)}} {d \over dk}
{\rm Re}~\Sigma^{\uparrow(\downarrow)} (k, \omega 
= \xi_k^{\uparrow(\downarrow)}) |_{k=k_F^{\uparrow(\downarrow)}}.
\end{equation}
The on-shell effective mass approximation given in
Eq.~(\ref{eq:onshell}) is known to be a better approximation {\em for
  the single-loop self-energy} calculation (compared, for example, to
solving the full Dyson's equation iteratively) as it is more
consistent with the leading-order nature of the self-energy
approximation itself-- in fact, the on-shell approximation of
Eq.~(\ref{eq:onshell}) is the natural spin-polarized generalization of
the quasiparticle effective mass renormalization in the usual
spin-unpolarized case~\cite{mass, rice}. Using the Feynman diagrams of
Fig.~\ref{fig:feynman}, we have calculated the on-shell effective mass
renormalization $m^*/m$ for both majority and minority spin carriers
in 2D and 3D electron systems as functions of the $r_s$ and the
$\zeta$ parameters, and below we present these results. For the sake
of conceptual clarity we present both 2D and 3D results on the same
footing without incorporating any finite quasi-2D width corrections,
which are straightforward to incorporate and will reduce the 2D
renormalization by factors of $1.2 - 2$, depending on the carrier
density and details of 2D confinement~\cite{mass}. We take the bare
electron-electron interaction to be the usual `$1/r$' Coulomb
interaction in 2D or 3D system, and the bare single-particle energy
dispersion to be parabolic.

\begin{figure}[htbp]
  \centering
  \includegraphics[width=2.5in]{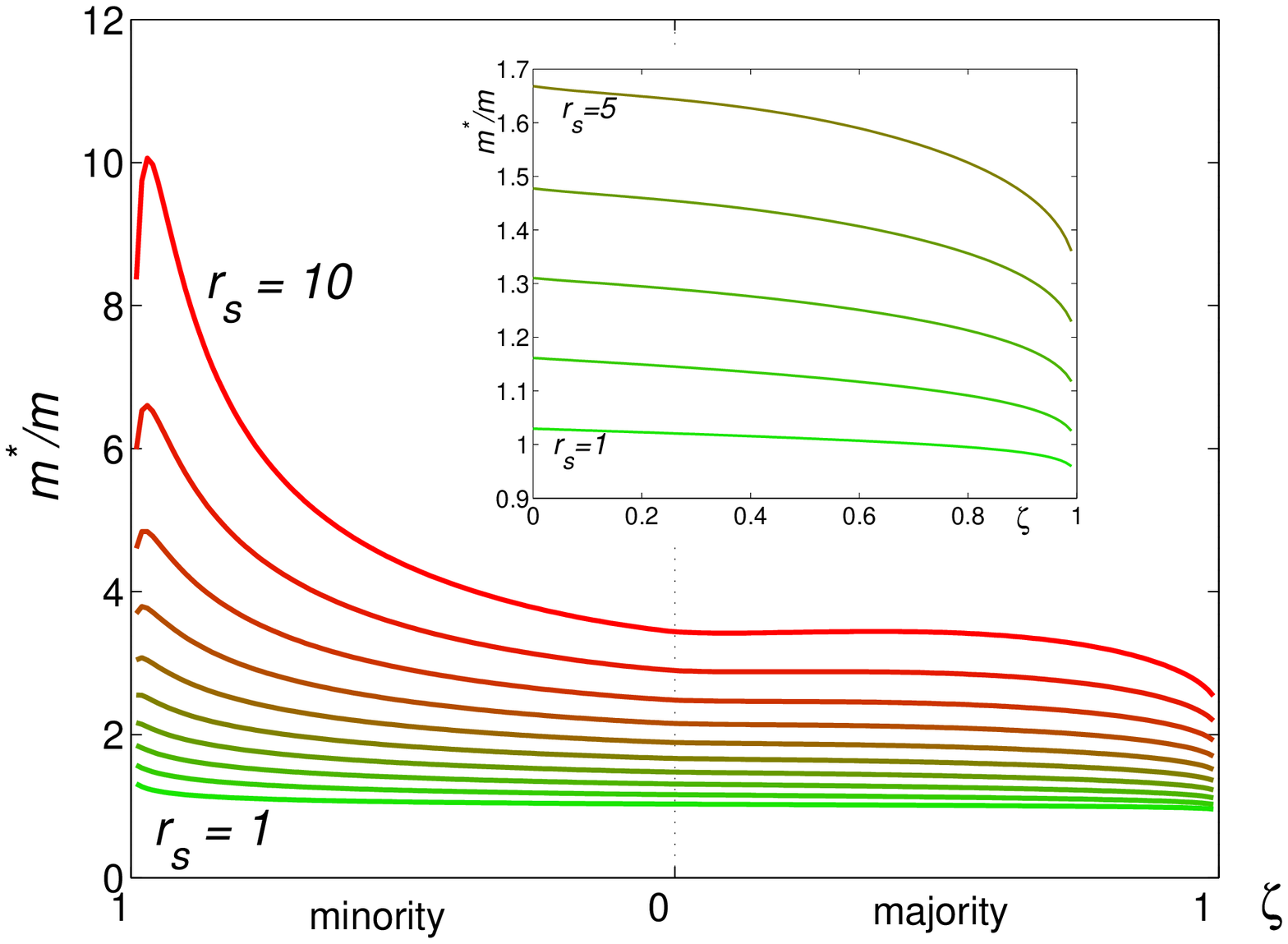}
  \caption{(Color online.) Calculated 2D effective mass as a function 
    of spin-polarization $\zeta$ at different $r_s$ values. Each curve
    corresponds to an $r_s$ value of $1$ to $10$ with increment $1$
    from bottom to top. Inset: zoom-in figure for majority effective
    mass as a function of $\zeta$ for $r_s = 1$ to $5$ from bottom to
    top with increment $1$.}
  \label{fig:2D}
\end{figure}

In Fig.~\ref{fig:2D} we present the calculated results for effective
mass in an ideal 2D electron system with Coulomb interaction. For the
majority electron mass, which is the likely experimentally measured
quantity, effective mass decreases with increasing spin polarization.
For small $\zeta$ values, this decrease is relatively small.  However
as $\zeta$ approaches unity (i.e. near full spin polarization), the
effect becomes much stronger. This is perhaps the reason for the
misconception held in some of the earlier literature that the
effective mass is spin-polarization independent, as it essentially is
near $\zeta = 0$, but certainly not for $\zeta \sim 1$.
Fig.~\ref{fig:2D} also shows that the minority effective mass
increases with $\zeta$ first, and as $\zeta$ approaches one, it
actually decreases sharply. The spin-polarization dependence of the
effective mass is thus quite non-trivial and non-monotonic.

\begin{figure}[htbp]
  \centering
  \includegraphics[width=2.5in]{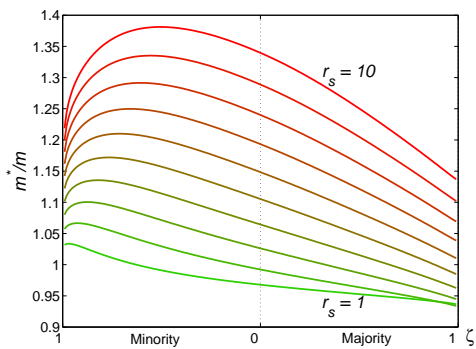}
  \caption{(Color online.) Calculated 3D effective mass as a function 
    of $\zeta$ for different $r_s$ values. Each curve corresponds to
    an $r_s$ value $1$ to $10$ with increment $1$ from bottom to top.}
  \label{fig:3D}
\end{figure}

In Fig.~\ref{fig:3D} we present the calculated results for the
effective mass in an ideal 3D electron system with Coulomb
interaction. The major difference between our 3D results shown in
Fig.~\ref{fig:3D} and 2D results shown in Fig.~\ref{fig:2D} is that
the 3D majority effective mass decreases more or less linearly with
increasing $\zeta$, but the non-monotonic dependence of the minority
effective mass on spin-polarization is manifestly present in 3D also,
except it is less sharp than in the 2D case.

\begin{figure}[htbp]
\centering \includegraphics[width=2.5in]{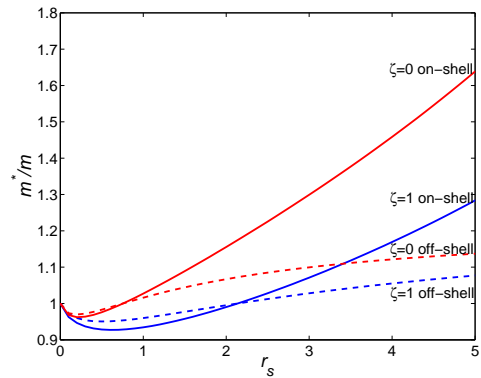}
  \caption{(Color online.) Calculated 2D majority effective mass as 
    a function of $r_s$ for on-shell and off-shell approximations.}
  \label{fig:onoff}
\end{figure}

In Fig.~\ref{fig:onoff} we show the 2D majority effective mass as a
function of $r_s$ at $\zeta = 0$ or $1$ for both on-shell and
off-shell approximations. In the off-shell approximation the full
Dyson's equation is solved for obtaining the quasiparticle effective
mass leading to $m^*_{\uparrow \downarrow} /m = \left[ 1 -
\partial_\omega \Sigma_{\uparrow \downarrow}(k, \omega) \right]/
\left[1+ (m/k) \partial_k \Sigma_{\uparrow \downarrow}(k, \omega)
\right]|_{k = k_{F \uparrow \downarrow}, \omega = 0}$. The important
difference between the definitions of the on-shell and the off-shell
effective mass leads to a large quantitative difference between their
calculated values, as apparent in Fig.~\ref{fig:onoff}. In general,
the many-body renormalization corrections are substantially suppressed
in the off-shell calculation ({\em within the single loop self-energy
  approximation}), and as has been argued extensively
elsewhere~\cite{rice, mass}, the on-shell approximation is the correct
effective mass approximation for the single-loop self-energy used in
our work, both for the sake of consistency and for the approximate
inclusion of vertex correction. We note that the spin-polarization
dependence of the off-shell effective mass (as well as the many-body
renormalization itself) is weaker than the on-shell result. We point
out that the only reason we are providing the off-shell effective mass
results in this paper (although within our approximation scheme of
Fig.~\ref{fig:feynman}, the on-shell mass is the appropriate one) is
to emphasize the fact that the off-shell result (which is often used
in the literature) is quantitatively highly inaccurate for the
one-loop self-energy approximation. For example, the on-shell 2D
effective mass in our theory agrees far better~\cite{mass} with
experiment than the corresponding off-shell results.

In further discussing our theoretical results for the
spin-polarization dependence of the quasiparticle effective mass
$m^*(r_s, \zeta)$ we first note that the $\zeta$-dependence is rather
weak in the high-density limit (small $r_s$). This trend can actually
be analytically demonstrated by obtaining the $\zeta$-dependence of
$m^*(r_s, \zeta)$ in the $r_s \to 0$ (and small $\zeta$) limit, which
for the single-loop self-energy gives through a straightforward
calculation:
\begin{equation}
\label{eq:2Dm}
{m^*_{\uparrow \downarrow} \over m} = 1 
+ {r_s \over \sqrt{2} \pi} \ln r_s
\mp {r_s \zeta \over \sqrt{2} \pi} \ln r_s,
\end{equation}
where $r_s \equiv m e^2 /\sqrt{\pi n}$ in 2D, and 
\begin{equation}
\label{eq:3Dm}
{m^*_{\uparrow \downarrow} \over m} = 1 
+ ({1 \over 2 \pi}) ({4 \over 9 \pi})^{1/3} \ln r_s
\mp ({1 \over 3 \pi}) ({4 \over 9 \pi})^{1/3} \zeta \ln r_s,
\end{equation}
where $r_s \equiv m e^2 (4 \pi n /3)^{-1/3}$ in 3D. Thus, in the small
$(r_s, \zeta)$ limit, the spin-polarization dependence of $m^*$ is
small and linear in the spin-polarization. For large
spin-polarization, however, the dependence of dynamical screening on
the spin-polarization is highly non-trivial, and the minority-spin
effective mass (note that the minority spin carrier density vanishes
as the spin-polarization approaches unity) shows a pronounced maximum
(both in 2D and 3D) which is not captured in the leading-order
asymptotic expansion in the small $r_s$ and $\zeta$ limit. A direct
observation of this non-monotonicity will indicate a non-trivial
many-body aspect of the spin polarization dependence of the spintronic
effective mass renormalization.

We can now re-examine the previously mentioned
experiments~\cite{vitkalov, shashkin, pudalov, tutuc, tan} using our
spintronic effective mass results. These experiments all use external
magnetic field induced Shubnikov-de Hass (SdH) oscillations to obtain
the effective mass of two-dimensional electron systems. The SdH
oscillation amplitude depends on the effective mass, temperature and
magnetic field according to the Dingle formula, through which
effective mass can be obtained by data fitting~\cite{vitkalov,
  shashkin, pudalov, tutuc, tan}. However, it is important to notice
that this is done in finite perpendicular magnetic field, which
invariably spin-polarizes the system. As we have already mentioned, it
is conceptually wrong to assume that the derived effective mass
corresponds to the zero-field mass because the effective mass depends
on the spin polarization. From our results shown in Fig.~\ref{fig:2D},
we see that when the polarization is less than $1/2$, effective mass
depends on the polarization weakly, and the above mentioned method of
experimentally determining the effective mass may serve as a good
approximation. However, as spin polarization approaches one (such as
the case in Ref.~\cite{shashkin}), the effective mass depends strongly
on the spin polarization, and fitting to the Dingle formula is no
longer a suitable method to obtain the finite field effective mass,
and certainly not the zero field effective mass. One must incorporate
the spin polarization dependence of the quasiparticle effective mass
in this situation in the experimental analysis.

\begin{figure}[htbp]
\centering \includegraphics[width=2.5in]{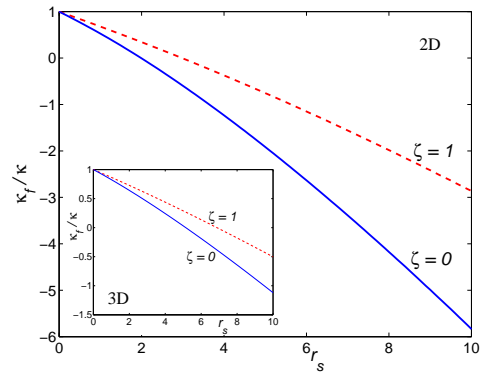}
  \caption{(Color Online.) Calculated 2D and 3D compressibility as
    a function of $r_s$ for fully spin-polarized and unpolarized
    states with $\kappa_f$ as the noninteracting unpolarized Fermi gas
    compressibility and $\kappa$ the renormalized compressibility.}
  \label{fig:comp}
\end{figure}

Since the single-particle self-energy, the density of states, the
dynamical screening, the Fermi momentum, and the Fermi energy are all
affected by spin-polarization, we expect all Fermi liquid parameters
(not just the effective mass) to be strongly dependent on the
spin-polarization parameter $\zeta$ (in addition to being dependent on
$r_s$). An important thermodynamic quantity is the system
compressibility, which is essentially the inverse of the volume
derivative of the pressure of the system.  As a related application of
the spin-polarization dependence of many-body effects we have
calculated the effect of finite spin-polarization on the
compressibility of the interacting spin-polarized 2D and 3D electron
systems within the same infinite ring-diagram approximation. We show
these results (only for the unpolarized and the fully polarized cases)
in Fig.~\ref{fig:comp}, where again many-body spin-polarization
corrections to the interacting compressibility are obvious. Since the
interacting compressibility in 2D electron systems can be directly
measured~\cite{eisenstein} with great accuracy, we suggest this as a
possible way of estimating the spin-polarization effect on the
many-body compressibility.

We conclude by emphasizing that the `spintronic' effective mass (and
compressibility) in spin-polarized carrier systems could be strongly
spin-polarization dependent and substantially different from the usual
unpolarized paramagnetic values. Such a spin-polarization dependence
should have serious implications in various spintronic applications.
For example, both spin Hall effect~\cite{kato, wunderlich} and spin
transistors~\cite{zutic2, datta} involve the carrier effective mass,
which would be explicitly spin-polarization dependent, complicating
understanding of the experimental data. We suggest that experiments be
carried out to directly test our predicted many-body spin-polarization
dependence of the carrier effective mass in 2D semiconductor
structure.

This work is supported by ONR, NSF, and LPS.


\end{document}